\documentclass[12pt]{article}
\usepackage{amsfonts}
\usepackage[dvips]{graphicx}
\usepackage{latexsym,amsmath,amssymb,graphics,stmaryrd}
\usepackage{cite}
\usepackage{array}
\usepackage{subfigure}

\usepackage{multirow}
\usepackage{epsfig}
\usepackage[dvips]{graphicx}
\usepackage[dvips]{color}
\usepackage{pstricks}
\usepackage{pst-node}
\usepackage{pst-plot}
\usepackage{dsfont}
\usepackage{amsthm}
\usepackage{graphics}
\usepackage{subfigure}
\usepackage{fancyhdr}
\usepackage[dvips]{color}
\usepackage{feynmp}

\fancypagestyle{plain}{
\fancyhf{}
\newcommand{\fpage}{\iffloatpage{}{\thepage}}
\fancyfoot[C]{\fpage}

}

\newcommand{\col}{~,}
\newcommand{\pnt}{~.}
\newcommand{\AdS}{\text{AdS}}

\newcommand{\YM}{\text{YM}}

\newcommand{\N}{\mathcal{N}}

\newcommand{\unitmatrix}{\mathds{1}}
\newcommand{\e}{\operatorname{e}}
\newcommand{\pfour}[4]{{}\boldsymbol{\{}#1,#2,#3,#4\boldsymbol{\}}{}}
\newcommand{\pthree}[3]{{}\boldsymbol{\{}#1,#2,#3\boldsymbol{\}}{}}
\newcommand{\ptwo}[2]{{}\boldsymbol{\{}#1,#2\boldsymbol{\}}{}}
\newcommand{\pone}[1]{{}\boldsymbol{\{}#1\boldsymbol{\}}{}}
\newcommand{\pid}{{}\boldsymbol{\{ \}}{}}
\newcommand{\dchi}[1]{\boldsymbol{\chi}(#1)}

\newlength{\neglength}
\newlength{\diameter}
    
\newcommand{\svertex}[2]{%
\fmfiequ{#1}{point length(#2)/2 of #2}
}
\newcommand{\dvertex}[3]{%
\fmfiequ{#1}{point length(#3)/3 of #3}
\fmfiequ{#2}{point 2length(#3)/3 of #3}
}
\newcommand{\vvertex}[3]{%
\fmfipath{px}
\fmfiequ{px}{(0,ypart(#2))..(100,ypart(#2))}
\fmfiequ{#1}{point xpart(#3 intersectiontimes px) of #3}
}

\newcommand{\beq}{\begin{equation}}
\newcommand{\eeq}{\end{equation}}
\newcommand{\bea}{\begin{eqnarray}}
\newcommand{\eea}{\end{eqnarray}}
\newcommand{\ena}{\end{eqnarray}}

\renewcommand{\b}{\beta}

\renewcommand{\d}{\delta}

\newcommand{\z}{\zeta}

\newcommand{\plainwrap}[4]{%
\fmfipath{pi[]}
\fmfiset{pi1}{vloc(__#1) ..controls (-0.175w,ypart(vloc(__#1))) and (-0.175w,-0.15w) .. (xpart(vloc(__#2)),-0.15w)}
\fmfiset{pi2}{(xpart(vloc(__#2)),-0.15w) ..(xpart(vloc(__#3)),-0.15w)}

\fmfiset{pi3}{(xpart(vloc(__#3)),-0.15w) ..controls (1.175w,-0.15w) and (1.175w,ypart(vloc(__#4))) .. vloc(__#4)}
\fmfi{plain}{pi1 ..pi2 ..pi3}
}
\newcommand{\wigglywrap}[4]{%
\fmfipath{pi[]}
\fmfiset{pi1}{#1 ..controls (-0.175w,ypart(#1)) and (-0.175w,-0.15w) .. (xpart(vloc(__#2)),-0.15w)}
\fmfiset{pi2}{(xpart(vloc(__#2)),-0.15w) ..(xpart(vloc(__#3)),-0.15w)}

\fmfiset{pi3}{(xpart(vloc(__#3)),-0.15w) ..controls (1.175w,-0.15w) and (1.175w,ypart(#4)) .. #4}
\fmfi{wiggly}{pi1 ..pi2 ..pi3}
}

\newcommand{\WoneplainB}{%
\fmftop{v1}
\fmfbottom{v4}
\fmfforce{(0.125w,h)}{v1}
\fmfforce{(0.125w,0)}{v4}
\fmffixed{(0.25w,0)}{v1,v2}
\fmffixed{(0.25w,0)}{v2,v3}
\fmffixed{(0.25w,0)}{v4,v5}
\fmffixed{(0.25w,0)}{v5,v6}
\fmf{plain,tension=0.5,right=0.25}{v1,vc1}
\fmf{plain,tension=0.5,left=0.25}{v2,vc1}
\fmf{plain}{vc1,vc2}
\fmf{plain,tension=0.5,left=0.25}{v4,vc2}
\fmf{plain,tension=0.5,right=0.25}{v5,vc2}
\fmf{plain}{v3,v6}
\fmf{plain,tension=0.5,right=0,width=1mm}{v4,v6}
\fmfposition
\fmfipath{p[]}
\fmfiset{p1}{vpath(__v1,__vc1)}
\fmfiset{p2}{vpath(__v2,__vc1)}
\fmfiset{p3}{vpath(__v3,__v6)}
\fmfiset{p4}{vpath(__vc1,__vc2)}
\fmfiset{p5}{vpath(__v4,__vc2)}
\fmfiset{p6}{vpath(__v5,__vc2)}

}

\newcommand{\WthreeplainB}{%
\fmftop{v1}
\fmfbottom{v4}
\fmfforce{(0.125w,h)}{v1}
\fmfforce{(0.125w,0)}{v4}
\fmffixed{(0.25w,0)}{v1,v2}
\fmffixed{(0.25w,0)}{v2,v3}
\fmffixed{(0.25w,0)}{v4,v5}
\fmffixed{(0.25w,0)}{v5,v6}
\fmffixed{(0,whatever)}{vc1,vc2}
\fmffixed{(0,whatever)}{vc3,vc4}
\fmf{plain,tension=0.5,right=0.25}{v2,vc1}
\fmf{plain,tension=0.5,left=0.25}{v3,vc1}
\fmf{plain,right=0.25}{v1,vc3}
\fmf{plain,tension=0.5,left=0.25}{v4,vc4}
\fmf{plain,tension=0.5,right=0.25}{v5,vc4}
\fmf{plain,right=0.25}{v6,vc2}
\fmf{plain,tension=1}{vc1,vc2}
\fmf{plain,tension=0.5}{vc2,vc3}
\fmf{plain,tension=1}{vc3,vc4}
\fmf{plain,tension=0.5,right=0,width=1mm}{v4,v6}
\fmfposition
\fmfipath{p[]}
\fmfiset{p1}{vpath(__v1,__vc3)}
\fmfiset{p2}{vpath(__vc3,__vc4)}
\fmfiset{p3}{vpath(__v4,__vc4)}
\fmfiset{p4}{vpath(__v3,__vc1)}
\fmfiset{p5}{vpath(__vc1,__vc2)}
\fmfiset{p6}{vpath(__v6,__vc2)}
}

\newcommand{\Wfourplain}{%
\fmftop{v1}
\fmfbottom{v5}
\fmfforce{(0.125w,h)}{v1}
\fmfforce{(0.125w,0)}{v5}
\fmffixed{(0.25w,0)}{v1,v2}
\fmffixed{(0.25w,0)}{v2,v3}
\fmffixed{(0.25w,0)}{v3,v4}
\fmffixed{(0.25w,0)}{v5,v6}
\fmffixed{(0.25w,0)}{v6,v7}
\fmffixed{(0.25w,0)}{v7,v8}
\fmffixed{(0,whatever)}{vc1,vc2}
\fmffixed{(0,whatever)}{vc3,vc4}
\fmffixed{(0,whatever)}{vc5,vc6}
\fmf{plain,tension=0.25,right=0.25}{v1,vc1}
\fmf{plain,tension=0.25,left=0.25}{v2,vc1}
\fmf{plain,left=0.25}{v5,vc2}
\fmf{plain,tension=1,left=0.25}{v3,vc3}
\fmf{plain,tension=1,left=0.25}{v4,vc5}
\fmf{plain,left=0.25}{v6,vc4}
\fmf{plain,tension=0.25,left=0.25}{v7,vc6}
\fmf{plain,tension=0.25,right=0.25}{v8,vc6}
\fmf{plain,tension=0.5}{vc1,vc2}
\fmf{plain,tension=0.5}{vc2,vc3}
\fmf{plain,tension=0.5}{vc3,vc4}
\fmf{plain,tension=0.5}{vc4,vc5}
\fmf{plain,tension=0.5}{vc5,vc6}
\fmf{plain,tension=0.5,right=0,width=1mm}{v5,v8}
\fmfposition
\fmfipath{p[]}
\fmfiset{p1}{vpath(__v1,__vc1)}
\fmfiset{p2}{vpath(__vc1,__vc2)}
\fmfiset{p3}{vpath(__v5,__vc2)}
\fmfiset{p4}{vpath(__v4,__vc5)}
\fmfiset{p5}{vpath(__vc5,__vc6)}
\fmfiset{p6}{vpath(__v8,__vc6)}
}

\newcommand{\WfourplainB}{%
\fmftop{v1}
\fmfbottom{v5}
\fmfforce{(0.125w,h)}{v1}
\fmfforce{(0.125w,0)}{v5}
\fmffixed{(0.25w,0)}{v1,v2}
\fmffixed{(0.25w,0)}{v2,v3}
\fmffixed{(0.25w,0)}{v3,v4}
\fmffixed{(0.25w,0)}{v4,v9}
\fmffixed{(0.25w,0)}{v5,v6}
\fmffixed{(0.25w,0)}{v6,v7}
\fmffixed{(0.25w,0)}{v7,v8}
\fmffixed{(0.25w,0)}{v8,v10}
\fmffixed{(0,whatever)}{vc1,vc2}
\fmffixed{(0,whatever)}{vc3,vc4}
\fmffixed{(0,whatever)}{vc5,vc6}
\fmffixed{(0,whatever)}{vc7,vc8}
\fmf{plain,tension=0.25,right=0.25}{v1,vc1}
\fmf{plain,tension=0.25,left=0.25}{v2,vc1}
\fmf{plain,left=0.25}{v5,vc2}
\fmf{plain,tension=1,left=0.25}{v3,vc3}
\fmf{plain,tension=1,left=0.25}{v4,vc5}
\fmf{plain,tension=1,left=0.25}{v9,vc7}
\fmf{plain,left=0.25}{v7,vc6}
\fmf{plain,tension=0.25,left=0.25}{v8,vc8}
\fmf{plain,tension=0.25,right=0.25}{v10,vc8}
\fmf{plain,left=0.25}{v6,vc4}
\fmf{plain,tension=0.5}{vc1,vc2}
\fmf{plain,tension=0.5}{vc2,vc3}
\fmf{plain,tension=0.5}{vc3,vc4}
\fmf{plain,tension=0.5}{vc4,vc5}
\fmf{plain,tension=0.5}{vc5,vc6}
\fmf{plain,tension=0.5}{vc6,vc7}
\fmf{plain,tension=0.5}{vc7,vc8}
\fmf{plain,tension=0.5,right=0,width=1mm}{v5,v10}
\fmfposition
\fmfipath{p[]}
\fmfiset{p1}{vpath(__v1,__vc1)}
\fmfiset{p2}{vpath(__vc1,__vc2)}
\fmfiset{p3}{vpath(__v5,__vc2)}
\fmfiset{p4}{vpath(__v4,__vc5)}
\fmfiset{p5}{vpath(__vc5,__vc6)}
\fmfiset{p6}{vpath(__v8,__vc6)}
}

\DeclareMathOperator{\tr}{tr}

\DeclareMathOperator{\perm}{P}
\DeclareMathOperator{\dperm}{\mathbf{P}}
\numberwithin{equation}{section}
\newlength{\eqoff}
\newlength{\eqofftwo}
\newlength{\unit}
\psset{xunit=\unit,yunit=\unit,runit=\unit}
\newlength{\linew}
\psset{linewidth=\linew}
\begin{document}
\begin{fmffile}{fullgraphs}

\fmfcmd{
wiggly_len := 2mm;
vardef wiggly expr p_arg =
 save wpp,len;
 numeric wpp,alen;
 wpp = ceiling (pixlen (p_arg, 10) / wiggly_len);
 len=length p_arg;
 for k=0 upto wpp - 1:
  point arctime k/(wpp-1)*arclength(p_arg) of p_arg of p_arg
    {direction arctime k/(wpp-1)*arclength(p_arg) of p_arg of p_arg rotated wiggly_slope} ..
  point  arctime (k+.5)/(wpp-1)*arclength(p_arg) of p_arg of p_arg
 {direction arctime (k+.5)/(wpp-1)*arclength(p_arg) of p_arg of p_arg rotated - wiggly_slope} ..
 endfor
 if cycle p_arg: cycle else: point infinity of p_arg fi
enddef;
}
\fmfcmd{%
marksize=2mm;
def draw_mark(expr p,a) =
  begingroup
    save t,tip,dma,dmb; pair tip,dma,dmb;
    t=arctime a of p;
    tip =marksize*unitvector direction t of p;
    dma =marksize*unitvector direction t of p rotated -45;
    dmb =marksize*unitvector direction t of p rotated 45;
    linejoin:=beveled;
    draw (-.5dma.. .5tip-- -.5dmb) shifted point t of p;
  endgroup
enddef;
style_def derplain expr p =
    save amid;
    amid=.5*arclength p;
    draw_mark(p, amid);
    draw p;
enddef;
def draw_point(expr p,a) =
  begingroup
    save t,tip,dma,dmb,dmo; pair tip,dma,dmb,dmo;
    t=arctime a of p;
    tip =marksize*unitvector direction t of p;
    dma =marksize*unitvector direction t of p rotated -45;
    dmb =marksize*unitvector direction t of p rotated 45;
    dmo =marksize*unitvector direction t of p rotated 90;
    linejoin:=beveled;
    draw (-.5dma.. .5tip-- -.5dmb) shifted point t of p withcolor 0white;
    draw (-.5dmo.. .5dmo) shifted point t of p;
  endgroup
enddef;
style_def derplainpt expr p =
    save amid;
    amid=.5*arclength p;
    draw_point(p, amid);
    draw p;
enddef;
style_def dblderplain expr p =
    save amidm;
    save amidp;
    amidm=.5*arclength p-0.5mm;
    amidp=.5*arclength p+0.5mm;
    draw_mark(p, amidm);
    draw_point(p, amidp);
    draw p;
enddef;
}

\begin{titlepage}
\begin{flushright}
IFUM-920-FT \\
\end{flushright}
\vspace{7ex}

\Large
\begin {center}     
{\bf
Finite-size effects
 in the superconformal $\beta$-deformed $\mathcal{N}=4$ SYM}
\end {center}

\renewcommand{\thefootnote}{\fnsymbol{footnote}}

\large
\vspace{1cm}
\centerline{F.\ Fiamberti ${}^{a,b}$, A.\ Santambrogio ${}^b$, 
C.\ Sieg ${}^b$, D.\ Zanon ${}^{a,b}$
\footnote[1]{\noindent \tt
francesco.fiamberti@mi.infn.it \\
\hspace*{6.3mm}alberto.santambrogio@mi.infn.it \\ 
\hspace*{6.3mm}csieg@mi.infn.it \\ 
\hspace*{6.3mm}daniela.zanon@mi.infn.it}}
\vspace{4ex}
\normalsize
\begin{center}
\emph{$^a$  Dipartimento di Fisica, Universit\`a degli Studi di Milano, \\
Via Celoria 16, 20133 Milano, Italy}\\
\vspace{0.2cm}
\emph{$^b$ INFN--Sezione di Milano,\\
Via Celoria 16, 20133 Milano, Italy}
\end{center}
\vspace{0.5cm}
\rm
\abstract
\normalsize 
\noindent We study finite size effects for composite operators in the $SU(2)$ sector of the  superconformal $\beta$-deformed
${\cal{N}}=4$ SYM theory. In particular we concentrate on the spectrum of one single magnon. Since in this theory one-impurity states are non BPS we compute their anomalous dimensions including wrapping contributions up to four loops and discuss higher order effects. 
\vfill
\end{titlepage} 

\section{Introduction}

Recently remarkable progress has been achieved  in studying integrability properties within the arena of the AdS/CFT correspondence \cite{maldacena:1998re}. 
The ideal setting for performing this analysis is provided by the planar limit of the
 superconformal $\mathcal{N}=4$ SYM theory and its string dual in the $\AdS_5\times\text{S}^5$ background.
 Indeed major efforts have been devoted to the comparison of the spectra on the two sides of the correspondence, i.e. the anomalous dimensions of gauge invariant operators in the gauge theory versus
 the mass spectra in the corresponding string sector. Higher loop perturbative calculations in the gauge theory have been performed taking advantage of the
 quantum spin chain description for which an Hamiltonian and a corresponding asymptotic  Bethe ansatz can be constructed \cite{Minahan:2002ve,Beisert:2003tq,Beisert:2003ys,Staudacher:2004tk,Beisert:2004hm}. In particular important results have been obtained for operators of infinite length, since in this case the dynamics simplifies considerably
and it is essentially encoded in an exact, factorized S-matrix corrected by a dressing phase~\cite{Arutyunov:2004vx,Hernandez:2006tk,Beisert:2006ib,Beisert:2006ez}.

In order to deepen our understanding it becomes crucial to take into account finite size effects.
On the string theory side there have been recent papers addressing this issue and studying finite size contributions in the spectrum of magnons~\cite{SchaferNameki:2006gk,SchaferNameki:2006ey,
Arutyunov:2006gs,Minahan:2008re,Heller:2008at,Ramadanovic:2008qd,Hatsuda:2008gd,Gromov:2008ie}. In \cite{Penedones:2008rv} wrapping effects in some toy models were studied.

On the field theory side, i.e. in the $\mathcal{N}=4$ SYM theory, finite size effects arise from wrapping interactions \cite{Beisert:2004hm,Sieg:2005kd}.  The simplest situation in which this kind of interactions shows up is at four loops in the anomalous dimensions of the composite operator  $ \tr(\phi[Z,\phi]Z)$.
We have performed this calculation~\cite{us,uslong} and found a new type of contributions proportional to $\z(5)$ that increases the order of transcendentality of the anomalous dimension\footnote{Following our paper a four-loop calculation, not in accordance with our result, was presented in \cite{Keeler:2008ce}. }. While this result is not in contradiction with the Kotikov-Lipatov transcendentality in the universal scaling function, it was not expected in previous conjectures for the anomalous dimension of the composite operator~\cite{Rej:2005qt,Beisert:2006ez,Kotikov:2007cy}. It would be nice to have an independent check of our result\footnote{Such a check is provided by the later result in \cite{Bajnok:2008bm}.}, but if one attempts higher order calculations the algebra becomes immediately too cumbersome to deal with. Alternatively one can study a
less symmetric theory that might exhibit new features to be compared to string theory known results. Such an example is provided by an exactly marginal deformation of 
${\cal N}=4$ SYM theory preserving ${\cal N}=1$ supersymmetry. The deformed theory is obtained modifying the original ${\cal N}=4$ superpotential for the
chiral superfields in the following way
\beq
ig\,\tr\left(\phi\,\psi\,Z -  \phi\,Z\,\psi\right)~\longrightarrow ~ih\,\tr\left(\e^{i\pi\b} \phi\,\psi\,Z - \e^{-i\pi\b} \phi\,Z\,\psi\right)
\label{deformation}
\eeq
where in general  $h$ and $\b$ are complex constants.
In \cite{Leigh:1995ep} it was argued that this $\b$-deformed ${\cal N}=1$
theory becomes conformally invariant,
i.e. the deformation becomes exactly marginal, if one condition is satisfied by
the constants $h$ and $\b$.  More precisely it has been shown that for a real deformation parameter  this theory becomes superconformal, in the planar limit to all perturbative orders~\cite{Mauri:2005pa}, if 
\beq
h\bar{h}=g^2\col
\label{betareal}
\eeq
where $g=g_\YM$ is the Yang-Mills coupling constant. We also define 
the 't Hooft coupling constant
\begin{equation}\label{lambdadef}
\lambda=\frac{g^2N}{16\pi^2}\pnt
\end{equation}

Via the AdS/CFT correspondence this  $\beta$-deformed ${\cal{N}}=4$ SYM theory is expected to be equivalent to the Lunin-Maldacena string theory background~\cite{Lunin:2005jy}. The existence of integrable structures in the deformed string background has been analyzed in~\cite{Frolov:2005dj,Frolov:2005ty}. 
Finite size effects of single magnons were discussed in \cite{Bykov:2008bj}.
Integrability of the deformed field theory was studied in 
\cite{Berenstein:2004ys,Beisert:2005if}.

In this paper we want to study the anomalous dimension of short 
operators in the superconformal deformed ${\cal{N}}=4$ SYM theory including wrapping contributions to be compared to finite size effects in the corresponding string theory.\\
The anomalous dimension for a composite operator $\mathcal{O}$ is extracted from the $1/\varepsilon$ pole of the graphs contributing to its renormalization: for an operator undergoing multiplicative renormalization it is defined as
\begin{equation}
\gamma(\mathcal{O})=\lim_{\varepsilon\rightarrow0}\left[\varepsilon g\frac{\mathrm{d} }{\mathrm{d} g}\log\mathcal{Z}_{\mathcal{O}}(g,\varepsilon)\right]\col
\end{equation}
where
\begin{equation*}
\mathcal{O}_{\mathrm{ren}}=\mathcal{Z}_\mathcal{O}\mathcal{O}_{\mathrm{bare}}\pnt
\end{equation*}
In presence of mixing among different operators, the second equation should be understood in matrix form and the first one is still valid for the eigenstates of the renormalization matrix.\\

\noindent Our strategy will be the following:

 We perform perturbative calculations using a superfield approach. All the superspace conventions, $D$-algebra techniques and shortcuts are explained in detail in~\cite{Gates:1983nr,uslong} and will not be repeated here.
 
We compute higher-loop integrals using the Gegenbauer polynomial $x$-space technique \cite{Chetyrkin:1980pr} which we will review and refine in a forthcoming publication \cite{usGPXT}. We will also make use repeatedly of the results already listed in \cite{uslong}.
  
 Here we focus primarily on the calculations of the anomalous dimensions of composite operators that exhibit relevant differences as compared to the $\mathcal{N}=4$ SYM case.
The major novelty of the deformed theory is given by the fact that one-impurity states are not protected by supersymmetry. The shortest\footnote{The length-two operator $\tr(\phi Z)$ was shown to be protected in \cite{Penati:2005hp,Freedman:2005cg}.} 
such a state is given by the length-three, single-impurity operator 
\beq
{\cal{O}}_{1,3}= \tr(\phi Z Z)
\pnt
\label{oneimpurity3}
\eeq
We will compute its anomalous dimensions up to three loops. 
In order to achieve this goal we proceed following the same lines of reasoning as in \cite{us,uslong }.

First we write the dilatation operator up to the third order for the $\b$-deformed theory. In this way we obtain the correct Hamiltonian only for 
operators in the asymptotic limit. \\
In order to derive the correct result when it is applied to 
a state of length three, we have to subtract the range four interactions and
add explicitly the wrapping contributions.\\
This is done in Sections \ref{sec:dilop} and \ref{sec:oneimpstate}.
(More precisely in Section \ref{sec:dilop} we compute the deformed dilatation operator up to fourth order since it will be useful later on).

Then using the same technique we compute the anomalous dimension of the length-four, single state operator
\beq
{\cal{O}}_{1,4}= \tr(\phi Z Z Z)
\label{oneimpurity4}
\eeq
up to four loops. This computation is presented in the first part of Section 
\ref{sec:fourloops}.

We turn to the length-four,  two-impurity operators 
\beq
\tr(\phi\phi Z Z)\col\qquad\qquad\tr(\phi Z\phi Z)
\label{twoimpurity}
\eeq
in the second part of Section \ref{sec:fourloops}, where we exploit the knowledge of the wrapping dilatation operator for the undeformed case \cite{uslong} to compute their anomalous dimensions.

Finally we analyze again the simplest single-impurity operator ${\cal{O}}_{1,L}= \tr(\phi Z^{L-1})$,  and we attempt to go to higher order $L$ in perturbation theory. Even if  exact calculations are too difficult and out of reach, still we are able to compute whole classes of diagrams that allow us to make some plausible conjectures. These are described and collected in the last section of the paper.


\section{Dilatation operator}
\label{sec:dilop}

As anticipated in the introduction in order to compute the anomalous dimensions of single-impurity operators it is convenient to make use  of the asymptotic dilatation operator. In fact this allows to avoid the explicit computation of a large number of diagrams.
Thus we need derive the expression of the asymptotic dilatation operator for the $\b$-deformed theory. We now show how this can be obtained from the knowledge of the Hamiltonian of the $\N=4$ theory.

First of all we recall the form of a standard permutation of fields at sites $i$ and $j$ which is given by
\begin{equation}
\perm_{i,j}=\frac{1}{2}\left[\unitmatrix_{i,j}+\vec{\sigma}_i\cdot\vec{\sigma}_j\right]=\frac{1}{2}\left[\unitmatrix_{i,j}+\sigma_i^3\sigma_j^3+\sigma_i^+\sigma_j^-+\sigma_i^-\sigma_j^+\right]
\pnt
\end{equation}
Using these permutations we can build a set of standard basis operators
\begin{equation}
\label{basisops}
\{a_1,\dots,a_n\}=\sum_{r=0}^{L-1}\perm_{a_1+r,\;a_1+r+1}\cdots \perm_{a_n+r,\;a_n+r+1}
\pnt
\end{equation}
The dilatation operator for the $\N=4$ theory can be written in terms of these operators.

We now look for a similar set of basis operators for the deformed theory. To this end we consider deformed permutations~\cite{Berenstein:2004ys}
\begin{equation}
\dperm_{i,j}=\frac{1}{2}\left[\unitmatrix_{i,j}+\sigma_i^3\sigma_j^3+q^2\,\sigma_i^+\sigma_j^-+\bar{q}^2\,\sigma_i^-\sigma_j^+\right] 
\col\qquad\qquad q\equiv \e^{i\pi\b}
\label{defperm}
\end{equation}
and define corresponding deformed basis operators:
\begin{equation}
\label{defbasisops}
\pthree{a_1}{\dots}{a_n}=\sum_{r=0}^{L-1}\dperm_{a_1+r,\;a_1+r+1}\cdots\dperm_{a_n+r,\;a_n+r+1}
\pnt
\end{equation}
Using  these definitions  the expansions of the chiral structures of Feynman diagrams in terms of basis operators have exactly the same coefficients as in the undeformed theory~\cite{uslong}. They are given by
\begin{equation}
\label{chiralstructs}
\begin{aligned}
\dchi{a,b,c,d}&=\pid-4\pone1
+\ptwo ab+\ptwo ac+\ptwo ad+\ptwo bc+\ptwo bd+\ptwo cd\\
&\phantom{{}={}}
-\pthree abc-\pthree abd-\pthree acd-\pthree bcd
+\pfour abcd
\col
\\
\dchi{a,b,c}&=-\pid+3\pone1
-\ptwo ab-\ptwo ac-\ptwo bc+\pthree abc
\col
\\
\dchi{a,b}&=\pid-2\pone1+\ptwo ab
\col
\\
\dchi{1}&=-\pid+\pone1\col\\
\dchi{}&=\pid
\pnt
\end{aligned}
\end{equation}
Clearly since  
\begin{equation}
\lim_{q,\bar{q}\rightarrow1}\pthree ab\dots=\{a,b,\dots\}
\end{equation}
all our deformed expressions reduce to the correct $\mathcal{N}=4$ expressions in the limit $q,\bar{q}\rightarrow1$.
Therefore the dependence on $q$ and $\bar{q}$ is  encoded entirely in the deformed basis operators~\eqref{defbasisops} and  we can look for the dilatation operator as a linear combination of operators~\eqref{defbasisops} with coefficients which are independent of $q$, $\bar{q}$.

Since in the limit $q,\bar{q}\rightarrow1$ the deformed  theory reduces correctly to the  $\mathcal{N}=4$ theory, the above observations allow us to conclude that the asymptotic dilatation operator of the deformed theory   is simply given by the corresponding dilatation operator of the $\mathcal{N}=4$ SYM theory~\cite{Beisert:2007hz} through the substitutions
\beq
\{a_1,\dots,a_n\}~\rightarrow~\pthree{a_1}{\dots}{a_n}
\pnt
\label{substitution}
\eeq
The deformed dilatation operator up to four loops is given explicitly in Table~\ref{dilatation}. We have verified that its eigenvalues agree with the solutions of the deformed Bethe equations~\cite{Beisert:2005if}.
\begin{table}
\begin{equation*}
\begin{aligned}
\boldsymbol{D_0}&=\dchi{}\\
\\\vspace{0.2cm}
\boldsymbol{D_1}&=-2\dchi{1}\\
\\\vspace{0.2cm}
\boldsymbol{D_2}
&=4\dchi{1}-2[\dchi{1,2}+\dchi{2,1}]\\
\\\vspace{0.2cm}
\boldsymbol{D_3}&=-24\dchi{1}
+16[\dchi{1,2}+\dchi{2,1}]
-4\dchi{1,3}\\
&\phantom{{}={}}
-4i\epsilon_2\dchi{1,3,2}
+4i\epsilon_2\dchi{2,1,3}
-4[\dchi{1,2,3}+\dchi{3,2,1}]\\
\\\vspace{0.2cm}
\boldsymbol{D_4}&={}+{}200\dchi{1}
-150[\dchi{1,2}+\dchi{2,1}]
+8(10+\epsilon_{3a})\dchi{1,3}
-4\dchi{1,4}\\
&\phantom{{}={}}
+60[\dchi{1,2,3}+\dchi{3,2,1}]\\
&\phantom{{}={}}
+(8+8\z(3)+4\epsilon_{3a}-4i\epsilon_{3b}+2i\epsilon_{3c}-4i\epsilon_{3d})
\dchi{1,3,2}\\
&\phantom{{}={}}
+(8+8\z(3)+4\epsilon_{3a}+4i\epsilon_{3b}-2i\epsilon_{3c}+4i\epsilon_{3d})
\dchi{2,1,3}\\
&\phantom{{}={}}
-(4+4i\epsilon_{3b}+2i\epsilon_{3c})[\dchi{1,2,4}+\dchi{1,4,3}]\\
&\phantom{{}={}}
-(4-4i\epsilon_{3b}-2i\epsilon_{3c})[\dchi{1,3,4}+\dchi{2,1,4}]\\
&\phantom{{}={}}-(12+8\z(3)+4\epsilon_{3a})\dchi{2,1,3,2}\\
&\phantom{{}={}}
+(18+4\epsilon_{3a})[\dchi{1,3,2,4}+\dchi{2,1,4,3}]\\
&\phantom{{}={}}
-(8+2\epsilon_{3a}+2i\epsilon_{3b})[\dchi{1,2,4,3}+\dchi{1,4,3,2}]\\
&\phantom{{}={}}
-(8+2\epsilon_{3a}-2i\epsilon_{3b})[\dchi{2,1,3,4}+\dchi{3,2,1,4}]\\
&\phantom{{}={}}
-10[\dchi{1,2,3,4}+\dchi{4,3,2,1}]
\end{aligned}
\end{equation*}
\caption{Dilatation operator for the $\beta$-deformed theory up to four loops.}
\label{dilatation}
\end{table}

The knowledge of the asymptotic dilatation operator is very useful since it allows to compute the anomalous dimensions of  long composite operators, more precisely operators with a length such  that wrapping interactions do not contribute. As emphasized above, in the $\b$-deformed theory single-impurity states of the SU(2) sector are not protected in general.  If the corresponding operator $\mathcal{O}_\text{as}$ is long enough to avoid wrapping interactions, the anomalous dimension for such a state at a given perturbative order  can be obtained from the dilatation operator or alternatively from the all-loop result~\cite{Mauri:2005pa}
\begin{equation}
\gamma(\mathcal{O}_\text{as})=-1+\sqrt{1+g^2\Big\vert q-\frac{1}{q}\Big\vert^2\frac{N}{4\pi^2}}=-1+\sqrt{1+4g^2\sin^2(\pi\beta)\frac{N}{4\pi^2}}
\pnt
\label{gammaas}
\end{equation}
In the next two sections we will use these results  for the computations of the anomalous dimensions of short one-impurity states where finite size effects become important.


\section{One-impurity state at three loops}
\label{sec:oneimpstate}

\begin{figure}[h]
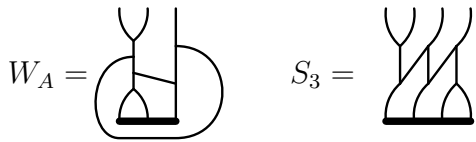

\unitlength=0.75mm
\settoheight{\eqoff}{$\times$}%
\setlength{\eqoff}{0.5\eqoff}%
\addtolength{\eqoff}{-12.5\unitlength}%
\settoheight{\eqofftwo}{$\times$}%
\setlength{\eqofftwo}{0.5\eqofftwo}%
\addtolength{\eqofftwo}{-7.5\unitlength}%
\begin{equation*}
\begin{aligned}
W_{A}=&
\raisebox{\eqoff}{%
\fmfframe(3,1)(1,4){%
\begin{fmfchar*}(20,20)
\fmftop{v1}
\fmfbottom{v4}
\fmfforce{(0.125w,h)}{v1}
\fmfforce{(0.125w,0)}{v4}
\fmffixed{(0.25w,0)}{v1,v2}
\fmffixed{(0.25w,0)}{v2,v3}
\fmffixed{(0.25w,0)}{v4,v5}
\fmffixed{(0.25w,0)}{v5,v6}
\fmffixed{(0,whatever)}{vc1,vc2}
\fmffixed{(0,whatever)}{vc2,vc3}
\fmffixed{(0,whatever)}{vc4,vc5}
\fmf{plain,tension=0.5,right=0.25}{v1,vc1}
\fmf{plain,tension=0.5,left=0.25}{v2,vc1}
\fmf{plain,tension=2}{vc1,vc7}
\fmf{plain,tension=2}{vc7,vc2}
\fmf{plain,tension=2}{vc2,vc3}
\fmf{plain,tension=0.5,left=0.25}{v4,vc3}
\fmf{plain,tension=0.5,right=0.25}{v5,vc3}
\fmf{plain,tension=0.5,left=0}{v3,vc4}
\fmf{plain,tension=0.5,right=0}{v6,vc5}
\fmf{plain,tension=0}{vc2,vc5}
\fmf{plain,tension=0.5}{vc4,vc5}
\fmfposition
\plainwrap{vc7}{v4}{v6}{vc4}
\fmf{plain,tension=0.5,right=0,width=1mm}{v4,v6}
\fmffreeze
\end{fmfchar*}}}
\;{\quad}&S_3=&
\raisebox{\eqoff}{%
\fmfframe(3,1)(1,4){%
\begin{fmfchar*}(20,20)
\Wfourplain
\end{fmfchar*}}}
\end{aligned}
\end{equation*}
\caption{Completely chiral diagrams (wrapping and subtraction)}
\label{diagrams-scalar}
\end{figure}

The three-loop contribution to the anomalous dimension of any asymptotic (i.e. of length greater than or equal to four) single-impurity operator $\mathcal{O}_\text{as}$ is given by
\begin{equation}
\label{3loop-asymptotic}
\gamma_3(\mathcal{O}_\text{as})=
256\,\lambda^3\sin^6(\pi\beta)
\pnt
\end{equation}
This is not a priori the correct value for the anomalous dimension of the length-three, single-impurity operator $\mathcal{O}_{1,3}=\tr(\phi Z Z)$ which is the shortest non-protected operator in the SU(2) sector.\\
Its exact anomalous dimension at three loops, taking wrapping interactions into account, can be obtained from the result for long states~\eqref{3loop-asymptotic} in two steps:
\begin{itemize}
\item subtract from~\eqref{3loop-asymptotic} the contributions of range four diagrams, which are not allowed for the length-three operator,
\item add the contributions of wrapping diagrams.
\end{itemize}
As explained in~\cite{uslong}, range four diagrams where one line is connected to the rest of the diagram by a single vector line sum up to zero. Therefore
the only relevant range four, single-impurity diagram is the one denoted by  $S_3$ in Figure~\ref{diagrams-scalar}. Its contribution is given by
\begin{equation}
S_3\rightarrow(g^2 N)^3\left(q-\bar{q}\right)^2\left(q^4+\bar{q}^4\right)I_0\sim-\frac{16}{3}\frac{\lambda^3}{\varepsilon}\sin^2(\pi\beta)\cos(4\pi\beta)
\col
\end{equation}
where $I_0$ is the momentum integral shown in Table~\ref{integrals}, the arrow denotes the result after $D$-algebra and the $\sim$ symbol means that we have kept only the $1/\varepsilon$ pole contribution. 
 The corresponding term we have to subtract from the asymptotic value~\eqref{3loop-asymptotic} is
\begin{equation}
\delta\gamma_3^{\,\text{s}}=-6\lim_{\varepsilon\rightarrow0}(\varepsilon S_3)=32\lambda^3\sin^2(\pi\beta)\cos(4\pi\beta)
\pnt
\end{equation}

Now we consider wrapping contributions. The various diagrams can be grouped into three classes according to their chiral structure. There is  one single diagram with only chiral lines, the one labeled $W_A$ in Figure~\ref{diagrams-scalar}. It gives
\begin{equation}
W_A\rightarrow(g^2 N)^3\left(q-\bar{q}\right)^2\left(q^4+\bar{q}^4\right)I_0\sim-\frac{16}{3}\frac{\lambda^3}{\varepsilon}\sin^2(\pi\beta)\cos(4\pi\beta)
\pnt
\end{equation}
Wrapping diagrams with chiral structures $\dchi{2,1}$ and $\dchi{1}$ are shown in the appendix, in Figures~\ref{diagrams12} and~\ref{diagrams1} respectively. 
Graphs with chiral structure $\dchi{1,2}$, which is the reflection of $\dchi{2,1}$, contribute exactly in the same manner.
After $D$-algebra we find  that the diagrams in each class sum up to zero separately. Therefore  these chiral structures do not contribute to the anomalous dimension. In the Tables~\ref{cancellations12} and~\ref{cancellations1} we show a possible way to combine these diagrams in pairs to have a full, explicit cancellation.

We conclude that the only wrapping contribution  to the anomalous dimension of $\mathcal{O}_{1,3}$ comes from $W_A$ and it is equal to
\begin{equation}
\delta\gamma_3^\text{w}=-6\lim_{\varepsilon\rightarrow0}(\varepsilon W_A)=32\lambda^3\sin^2(\pi\beta)\cos(4\pi\beta)
\pnt
\end{equation}

The correct three-loop anomalous dimension of $\mathcal{O}_{1,3}$ is then given by
\begin{equation}
\gamma_3(\mathcal{O}_{1,3})=\gamma_3(\mathcal{O}_\text{as})-\delta\gamma_3^{\,\text{s}}+\delta\gamma_3^\text{w}=256\,\lambda^3\sin^6(\pi\beta)
\pnt
\end{equation}
Since $\delta\gamma_3^{\,\text{s}}=\delta\gamma_3^\text{w}$  the final result happens to be equal to the asymptotic one in \eqref{3loop-asymptotic}. This cancellation is very likely to be a peculiarity of the three loop calculation and we expect it would no longer hold at four loops.
 In fact as we will see in the next section the four-loop anomalous dimension of the length-four $\mathcal{O}_{1,4}$ operator is different from its asymptotic value.\\
It would be nice to perform the four-loop calculation for $\mathcal{O}_{1,3}$ but at the fourth order one has to deal with the insurgence of the three-vector vertex which increases dramatically  the number and the complexity of the relevant diagrams to be computed.

\section{Four loops}
\label{sec:fourloops}

In this section we compute the anomalous dimension of one- and two-impurity states at four loops. In \cite{uslong} we wrote down the dilatation operator comprehensive of wrapping contributions on the length-four sector. This operator can be easily deformed as we have done for the asymptotic case in Section \ref{sec:dilop}. It can be used to obtain the anomalous dimensions both for one- and two-impurity states. 
However, in the single-impurity case, we chose to do an explicit calculation as in the three-loop case, to check our results and to avoid to use the integrability hypothesis which is hidden in the determination of the four-loop asymptotic dilatation operator.\footnote{We use \eqref{gammaas} which does not rely on 
integrability.}

\subsection{Single-impurity state}
Here we apply the same technique presented in the previous section to compute the exact anomalous dimension of the length-four, single-impurity state $\mathcal{O}_{1,4}$ at four loops. Again we start from the asymptotic result, valid for any single-impurity state $\mathcal{O}_\text{as}$ of length greater than four:
\begin{equation}
\label{4loop-asymptotic}
\gamma_4(\mathcal{O}_\text{as})=-2560\,\lambda^4\sin^8(\pi\beta)
\pnt
\end{equation}

As in the previous case one single diagram must be subtracted in order to get rid of the range five contributions. It contains only chiral interactions and it is shown in Figure~\ref{4loopSub}.
\begin{figure}[h]
\unitlength=0.75mm
\settoheight{\eqoff}{$\times$}%
\setlength{\eqoff}{0.5\eqoff}%
\addtolength{\eqoff}{-12.5\unitlength}%
\settoheight{\eqofftwo}{$\times$}%
\setlength{\eqofftwo}{0.5\eqofftwo}%
\addtolength{\eqofftwo}{-7.5\unitlength}%
\begin{equation}
\begin{aligned}
S_4=&
\raisebox{\eqoff}{%
\fmfframe(3,1)(1,4){%
\begin{fmfchar*}(20,20)
\WfourplainB
\fmfipair{w[]}
\fmfipair{wd[]}
\end{fmfchar*}}}
\end{aligned}
\end{equation}
\caption{Range-five diagram}
\label{4loopSub}
\end{figure}

\noindent Its contribution is given by
\begin{equation}
\begin{aligned}
S_4&\rightarrow
(g^2 N)^4 I_1\,[\dchi{1,2,3,4}+
\dchi{4,3,2,1}
]\\
&\sim\frac{5}{4}\frac{\lambda^4}{\varepsilon}\left[(\bar{q}^8+q^8)-2(\bar{q}^6+q^6)+(\bar{q}^4+q^4)\right]\\
&=
\frac{5}{2}\frac{\lambda^4}{\varepsilon}\left[\cos(8\pi\beta)-2\cos(6\pi\beta)+\cos(4\pi\beta)\right]
\pnt
\end{aligned}
\end{equation}
Thus the term to be subtracted from~\eqref{4loop-asymptotic} is
\begin{equation}
\delta\gamma_4^{\,\text{s}}=-8\lim_{\varepsilon\rightarrow0}(\varepsilon S_4)=-20\lambda^4\left[\cos(8\pi\beta)-2\cos(6\pi\beta)+\cos(4\pi\beta)\right]
\pnt
\end{equation}

The relevant wrapping diagrams are given by a subset of those contributing to the two impurity case which are listed in Figures 2 and C.1 to C.7 of~\cite{uslong}. In particular we need consider the diagram $W_{A2}$ and all the ones belonging to the classes $W_{B**}$, $W_{E**}$ and $W_{G**}$. The total contribution from each class can be read from Table~\ref{results-1}. 

\begin{table}[p]
\begin{center}
\small
\begin{tabular}{|ll|}
\hline
\multicolumn{2}{|c|}{}\\[-1ex]
\multicolumn{2}{|c|}{
$W_{A2\phantom{0}}\ \rightarrow\ (g^2 N)^4 I_2\, 
[\dchi{1,4,3,2}+\dchi{4,1,2,3}]
$ 
} \\[1ex]
\hline
\hline
& \\[-2ex]
$W_{B1\phantom{0}}\ \rightarrow\ $ & $-(g^2 N)^4 (I_3+I_4+2I_5) \,[\dchi{1,2,3}+\dchi{3,2,1}] $\\
$W_{B2\phantom{0}}\ \rightarrow\ $ & $(g^2 N)^4 I_3 \,[\dchi{1,2,3}+\dchi{3,2,1}]$\\
$W_{B3\phantom{0}}\ \rightarrow\ $ & $-(g^2 N)^4 I_2 \,[\dchi{1,2,3}+\dchi{3,2,1}]$\\
$W_{B4\phantom{0}}\ \rightarrow\ $ & $(g^2 N)^4 (I_2+I_4+2I_6) \,[\dchi{1,2,3}+\dchi{3,2,1}]$\\
[1ex]
\hline
\multicolumn{2}{|c|}{}\\[-1ex]
\multicolumn{2}{|c|}{
$\sum W_{B**}\  \rightarrow\ -2(g^2 N)^4 (I_5-I_6)\,[\dchi{1,2,3}+\dchi{3,2,1}]$
} \\[1ex]
\hline
\hline
& \\[-2ex]
$W_{E2\phantom{0}}\ \rightarrow\ $ & $-(g^2 N)^4 (I_2+I_4+2I_6) \,[\dchi{1,2}+\dchi{2,1}]$\\
$W_{E5\phantom{0}}\ \rightarrow\ $ & $(g^2 N)^4 I_2 \,[\dchi{1,2}+\dchi{2,1}]$\\
$W_{E11}\ \rightarrow\ $ & $(g^2 N)^4 (I_3+I_4+2I_5) \,[\dchi{1,2}+\dchi{2,1}]$\\
$W_{E14}\ \rightarrow\ $ & $-(g^2 N)^4 I_3 \,[\dchi{1,2}+\dchi{2,1}]$\\
[1ex]
\hline
\multicolumn{2}{|c|}{}\\[-1ex]
\multicolumn{2}{|c|}{
$\sum W_{E**}\ \rightarrow\ 2(g^2 N)^4 (I_5-I_6)\,[\dchi{1,2}+\dchi{2,1}]$
} \\[1ex]
\hline
\hline
& \\[-2ex]
$W_{G11}\ \rightarrow\ $ & $ 2(g^2 N)^4 I_1\,\dchi{1}$\\
$W_{G29}\ \rightarrow\ $ & $ -2(g^2 N)^4 I_{2}\,\dchi{1}$\\
[1ex]
\hline
\multicolumn{2}{|c|}{}\\[-1ex]
\multicolumn{2}{|c|}{
$\sum W_{G**}\  \rightarrow\  2(g^2 N)^4 (I_1-I_2)\,\dchi{1}$
} \\[1ex]
\hline
\end{tabular}
\end{center}
\caption{Four-loop wrapping contributions for the single impurity case}
\label{results-1}
\end{table}

Now we need the explicit expressions for the chiral structures $\dchi{\ldots}$ on single-impurity states. For the diagram $W_{A2}$ we can deform the corresponding wrapping structure $\chi(1,4,3,2) + \chi(4,1,2,3)$ described in \cite{uslong} or we can simply obtain the explicit expression from the diagram. We find
\begin{equation}
\dchi{1,4,3,2}+\dchi{4,1,2,3}
=\left[(\bar{q}^8+q^8)-2(\bar{q}^6+q^6)+(\bar{q}^4+q^4)\right]
\pnt
\end{equation}
All the other structures can be derived directly from the diagrams, as for $W_{A2}$, or computed using the general rules~\eqref{chiralstructs} and the definition of the deformed basis operators~\eqref{defbasisops}. We find
\begin{align}
W_{A2}&\sim
\left[\frac{5}{4}-\zeta(3)\right]\frac{\lambda^4}{\varepsilon}\left[(\bar{q}^8+q^8)-2(\bar{q}^6+q^6)+(\bar{q}^4+q^4)\right]\nonumber\\
&=
\left[\frac{5}{4}-\zeta(3)\right]\frac{\lambda^4}{\varepsilon}2\left[\cos(8\pi\beta)-2\cos(6\pi\beta)+\cos(4\pi\beta)\right]
\col\\
\sum W_{B**}
&\sim
\left[3\zeta(3)-5\zeta(5)\right]\frac{\lambda^4}{\varepsilon}\left[(\bar{q}^6+q^6)-2(\bar{q}^4+q^4)+(\bar{q}^2+q^2)\right]\nonumber\\
&=
\left[3\zeta(3)-5\zeta(5)\right]\frac{\lambda^4}{\varepsilon}2\left[\cos(6\pi\beta)-2\cos(4\pi\beta)+\cos(2\pi\beta)\right]
\col\\
\sum W_{E**}
&\sim
\left[-3\zeta(3)+5\zeta(5)\right]\frac{\lambda^4}{\varepsilon}\left[(\bar{q}^4+q^4)-2(\bar{q}^2+q^2)+2\right]\nonumber\\
&=
2\left[-3\zeta(3)+5\zeta(5)\right]\frac{\lambda^4}{\varepsilon}\left[\cos(4\pi\beta)-2\cos(2\pi\beta)+1\right]
\col\\
\sum W_{G**}
&\sim
2\zeta(3)\frac{\lambda^4}{\varepsilon}\left[(\bar{q}^2+q^2)-2\right]\nonumber\\
&=
4\zeta(3)\frac{\lambda^4}{\varepsilon}\left[\cos(2\pi\beta)-1\right]
\pnt
\end{align}
Therefore the wrapping contribution to the anomalous dimension is given by
\begin{equation}
\delta\gamma_4^\text{w}=-8\lim_{\varepsilon\rightarrow0}[\varepsilon(W_{A2}+\sum W_{B**}+\sum W_{E**}+\sum W_{G**})]
\pnt
\end{equation}
Finally collecting all the contributions we can write the exact anomalous dimension  of $\mathcal{O}_{1,4}$ at four loops

\begin{equation}
\begin{aligned}
\label{dim14}
\gamma_4(\mathcal{O}_{1,4})&=\gamma_4(\mathcal{O}_\text{as})-\delta\gamma_4^{\,\text{s}}+\delta\gamma_4^\text{w}\\
&=-16
\lambda^4
\big[160\sin^8(\pi\beta)-\zeta(3)\cos(8\pi\beta)+5\big(\zeta(3)-\zeta(5)\big)\cos(6\pi\beta)\\
&\phantom{{}={}-16\lambda^4\big[{}}-\big(10\zeta(3)-15\zeta(5)\big)\cos(4\pi\beta)+\big(11\zeta(3)-15\zeta(5)\big)\cos(2\pi\beta)\\
&\phantom{{}={}-16\lambda^4\big[{}}-5\big(\zeta(3)-\zeta(5)\big)\big]
\pnt
\end{aligned}
\end{equation}

\subsection{Two-impurity states}
In this subsection we consider the length-four,  two-impurity operators 
\beq
\label{24}
\tr(\phi\phi Z Z)\col\qquad\qquad\tr(\phi Z\phi Z)
\col
\eeq
and our aim is to compute their anomalous dimensions at four loops. 

We make use of the dilatation operator including wrapping interactions that we determined in \cite{uslong}, by adapting it to the deformed case through the substitution of the undeformed chiral structures with the ones given in \eqref{chiralstructs}
\begin{equation}
\begin{aligned}
\label{d4full}
\boldsymbol{D_4^\text{sub}}+\boldsymbol{\d D_4^\text{w}}
&=(200-16\zeta(3))\dchi{1}-(150-24\zeta(3)+40\zeta(5))[\dchi{1,2}+\dchi{2,1}]\\
&\phantom{{}={}}
+(88+8\epsilon_{3a}+24\zeta(3)-40\zeta(5))\dchi{1,3}\\
&\phantom{{}={}}+(60-24\zeta(3)+40\zeta(5))[\dchi{1,2,3}+\dchi{3,2,1}]\\
&\phantom{{}={}}-\Big(\frac{4}{3}-8\zeta(3)-4\epsilon_{3a}+4i\epsilon_{3b}-2i\epsilon_{3c}+4i\epsilon_{3d}\Big)\dchi{1,3,2}\\
&\phantom{{}={}}-\Big(\frac{20}{3}-8\zeta(3)-4\epsilon_{3a}-4i\epsilon_{3b}+2i\epsilon_{3c}-4i\epsilon_{3d}\Big)\dchi{2,1,3}\\
&\phantom{{}={}}+4(1-\zeta(3))\dchi{2,4,1,3}
-(10-8\zeta(3))[\dchi{1,4,3,2}+\dchi{4,1,2,3}]\\
&\phantom{{}={}}-(12+8\zeta(3)+4\epsilon_{3a})\dchi{2,1,3,2}
+(4-8\zeta(3))\dchi{4,1,3,2}
\col
\end{aligned}
\end{equation}
where $\boldsymbol{D_4^\text{sub}}$ contains the interactions with range up to four, while $\boldsymbol{\d D_4^\text{w}}$ contains the wrapping contributions \cite{uslong}.

Let us first notice that applying this operator to the single-impurity state $\mathcal{O}_{1,4}$ we immediately recover the four-loop anomalous dimension explicitly computed in the previous subsection and given in \eqref{dim14}.

In order to compute the anomalous dimensions of the two-impurity operators \eqref{24} we have to consider the full dilatation operator up to four loops:
\begin{equation}
\boldsymbol{D}=\boldsymbol{D_0}+\lambda\boldsymbol{D_1}+\lambda^2\boldsymbol{D_2}+\lambda^3\boldsymbol{D_3}+\lambda^4
\left(\boldsymbol{D_4^{\mathrm{sub}}}+\boldsymbol{\d D_4^{\mathrm{w}}}\right)
\pnt
\end{equation}
The application of this operator on the states \eqref{24} produces a mixing $2\times 2$ matrix whose eigenvalues are the anomalous dimensions we are looking for. We can write them as
\begin{equation}
\gamma^{(\pm)}=4+\lambda\gamma_1^{(\pm)}+\lambda^2\gamma_2^{(\pm)}+\lambda^3\gamma_3^{(\pm)}+\lambda^4\gamma_4^{(\pm)}
\pnt
\end{equation}
Finally introducing the definition
\begin{equation}
\Delta(\beta)=\frac{\sqrt{5+4\cos(4\pi\beta)}}{3}
\col
\end{equation}
we obtain the following results\footnote{After the appearance of \cite{Bajnok:2008bm} we corrected the rational part of this result.}
\begin{equation}
\begin{aligned}
\gamma_1^{(\pm)}&=6(1\mp\Delta(\beta))\col\\
\gamma_2^{(\pm)}&={}-{}3\big(5+3\Delta(\beta)^2\big)\pm\frac{3}{\Delta(\beta)}\big(1+7\Delta(\beta)^2\big)\col\\
\gamma_3^{(\pm)}&=6\big(19+9\Delta(\beta)^2\big)\pm\frac{3}{4\Delta(\beta)^3}\big(1-51\Delta(\beta)^2-165\Delta(\beta)^4-9\Delta(\beta)^6\big)\col\\
\gamma_4^{(\pm)}&={}-{}3\big(410-99\zeta(3)+120\zeta(5)\big)
-18\Delta(\beta)^2\big(10-13\zeta(3)+20\zeta(5)\big)\\
&\phantom{{}={}}
+81\Delta(\beta)^4\big(2-3\zeta(3)\big)\\
&\phantom{{}={}}
\pm\frac{3}{8\Delta(\beta)^5}\big[1-44\Delta(\beta)^2+6\Delta(\beta)^4\big(189+4\zeta(3)\big)\\
&\hphantom{{}={}\pm\frac{1}{8\Delta(\beta)^5}\big[{}}
+4\Delta(\beta)^6\big(809-468\zeta(3)+480\zeta(5)\big)
-27\Delta(\beta)^8\big(37-40\zeta(3)\big)\big]\col
\end{aligned}
\end{equation}
where the eigenstate with eigenvalue $\gamma^{(+)}$ becomes protected in the
undeformed theory with $\Delta(\beta=0)=1$.

We notice that in the deformed theory no BPS state survives in the $SU(2)$ sector. Thus, unlike in $\N=4$, the eigenstates of the dilatation operator change with the loop order.

\section{One-impurity states at higher orders}
\label{sec:higherorder}

Now we study once again the simplest one-impurity operators and  attempt to go beyond four loops. More specifically we look at the operator
 ${\cal{O}}_{1,L}= \tr(\phi Z^{L-1})$ and analyze its anomalous dimension at higher order $L$ in perturbation theory.
 Following our general strategy one would have to consider the asymptotic contributions from $\boldsymbol{D_L}$. Then one would have to compute 
$\boldsymbol{\d D^{\text{\bf s}}_L}$ in order to subtract the range $L+1$ interactions. Finally one would need all the  wrapping contributions $\boldsymbol{\d D^\text{\bf w}_L}$.\\
 The first step is actually simple since we do not need the knowledge of  
the asymptotic $\boldsymbol{D_L}$: the asymptotic contribution to the anomalous dimension of the single-impurity operator can be obtained directly from (\ref{gammaas}).\\
 Next we have to subtract the range $L+1$ interactions. As discussed in the previous sections there is only one diagram to be subtracted, i.e. the range $(L+1)$ graph  denoted by $S_L$ with the chiral structure $\dchi{1,2,\ldots,L}$.\\
 Then we have to consider the wrapping diagrams.
 At $L$ loops there will be wrapping contributions from one single diagram $W_{L,0}$ with only chiral interactions and from the classes with chiral structures
\begin{equation}
\begin{aligned}
W_{L,L-1}&:\qquad\dchi1\  +\  (L-1)\ \mathrm{ vectors}\col\\
W_{L,L-2}&:\qquad\dchi{1,2}\  +\  (L-2)\ \mathrm{ vectors}\col\\
&\qquad\qquad\qquad\qquad\vdots\\
W_{L,1}&:\qquad\dchi{1,2,\ldots,L-1}\  +\  1\ \mathrm{ vector}\pnt
\end{aligned}
\end{equation}

The general form of these contributions after $D$-algebra can be easily found for the structures with two, one and no vectors, for $L\geq4$. At four loops we have the complete result \begin{itemize}
\item $L=4$:
\begin{equation}
\begin{aligned}
W_{4,0}-S_4&=C_{4,0}\frac{1}{\varepsilon}\z(3)\col\\
W_{4,1}&=C_{4,1}\frac{1}{\varepsilon}[3\z(3)-5\z(5)]\col\\
W_{4,2}&=C_{4,2}\frac{1}{\varepsilon}[3\z(3)-5\z(5)]\col\\
W_{4,3}&=C_{4,3}\frac{1}{\varepsilon}\z(3)\col
\end{aligned}
\end{equation}
where the $C_{L,i}$ are rational prefactors.\\
Already at five loops wrapping diagrams with three vectors proliferate considerably and we have not embarked in their computations.  We computed the momentum integrals corresponding to the classes with two, one and no vectors for $L=5$ and 6, and the integrals for the cases of one and no vectors for $L=7$. The results read
\item $L=5$:
\begin{equation}
\begin{aligned}
W_{5,0}-S_5&=C_{5,0}\frac{1}{\varepsilon}\z(5)\col\\
W_{5,1}&=C_{5,1}\frac{1}{\varepsilon}[4\z(5)-7\z(7)]\col\\
W_{5,2}&=0\col
\end{aligned}
\end{equation}
\item $L=6$:
\begin{equation}
\begin{aligned}
W_{6,0}-S_6&=C_{6,0}\frac{1}{\varepsilon}[4\z(5)+35\z(7)]\col\\
W_{6,1}&=C_{6,1}\frac{1}{\varepsilon}\left[20\z(5)+49\z(7)-126\z(9)\right]\col\\
W_{6,2}&=C_{6,2}\frac{1}{\varepsilon}[10\z(5)-7\z(7)]\col
\end{aligned}
\end{equation}
\item $L=7$:
\begin{equation}
\begin{aligned}
W_{7,0}-S_7&=C_{7,0}\frac{1}{\varepsilon}[\z(7)+6\z(9)]\col\\
W_{7,1}&=C_{7,1}\frac{1}{\varepsilon}\left[2\z(7)+4\z(9)-11\z(11)\right]\pnt
\end{aligned}
\end{equation}
\end{itemize}

Even with the partial results we have listed above, several comments are in order. 

\noindent First of all, as a general observation, we recall that the anomalous dimensions of single-impurity states are not affected by the presence of a dressing phase. Therefore the transcendentality that we read in the results at the various loop orders is to be ascribed completely to finite size effects. 

We can summarize our findings as follows:\\
We have analyzed the anomalous dimensions of single-impurity states ${\cal{O}}_{1,L}= \tr(\phi Z^{L-1})$ at critical order, i.e. with operators with length equal to the loop order.\\
At order $L=3$ we have found that subtraction and wrapping contributions cancel each other and the net contribution to the anomalous dimension is the same as from its asymptotic value. Wrapping at three loops seems to be irrelevant.\\
At order $L=4$ we have computed exactly all the wrapping contributions to the anomalous dimension of
${\cal{O}}_{1,4}$ and found that the result contains terms proportional to $\z(3)$ and $\z(5)$.\\
Beyond four loops we have only partial results but a clear pattern seems to emerge: at every loop order the level of transcendentality is increased by two as compared to the previous order and no new rational part arises.

It becomes natural to compare this behavior to the one of the dressing phase: at three loop nothing happens, at four loop a contribution proportional to $\z(3) $ arises, at five loops $\z(5)$  appears and so on. What we have found indicates that
finite size effects resemble the behavior of the dressing phase contributions at the various loop orders~\cite{Eden:2006rx,Beisert:2006ez}, increasing by two the level of transcendentality.

Our calculation of the anomalous dimensions for the two-impurity state, where both the dressing phase and the wrapping contribute, confirms the above interpretation. 

We hope that these results might shed some light on how to implement the wrapping interactions into a modified Bethe ansatz.

Several other issues are still completely open: among them we mention the fact that it would be important to
compute finite size effects beyond critical order, i.e. compute the anomalous dimension of ${\cal{O}}_{1,L}$  at order $L+1$. Needless to say that now the major challenge resides in the
comparison of the finite size contributions  we have found in the weak coupling regime with the corresponding strong limit results from string theory computations.

\medskip

\section*{Acknowledgements}
\noindent This work has been supported in
part by INFN  and the European Commission RTN
program MRTN--CT--2004--005104.

\newpage

\renewcommand{\thefigure}{A.\arabic{figure}}
\setcounter{figure}{0}
\renewcommand{\thetable}{A.\arabic{table}}
\setcounter{table}{0}

\section*{Appendix}
\label{app}

\vspace{2cm}

\begin{figure}[h]
\unitlength=0.75mm
\settoheight{\eqoff}{$\times$}%
\setlength{\eqoff}{0.5\eqoff}%
\addtolength{\eqoff}{-12.5\unitlength}%
\settoheight{\eqofftwo}{$\times$}%
\setlength{\eqofftwo}{0.5\eqofftwo}%
\addtolength{\eqofftwo}{-7.5\unitlength}%
\begin{equation*}
\begin{aligned}
W_{B1}&=
\raisebox{\eqoff}{%
\fmfframe(3,1)(1,4){%
\begin{fmfchar*}(20,20)
\WthreeplainB
\fmfipair{w[]}
\fmfipair{wd[]}
\svertex{w3}{p3}
\svertex{w6}{p6}
\wigglywrap{w3}{v4}{v5}{w6}
\end{fmfchar*}}}
\;{\quad}&W_{B2}&=
\raisebox{\eqoff}{%
\fmfframe(3,1)(1,4){%
\begin{fmfchar*}(20,20)
\WthreeplainB
\fmfipair{w[]}
\fmfipair{wd[]}
\svertex{w5}{p5}
\svertex{w3}{p3}
\wigglywrap{w3}{v4}{v5}{w5}
\end{fmfchar*}}}
\;{\quad}&W_{B3}&=
\raisebox{\eqoff}{%
\fmfframe(3,1)(1,4){%
\begin{fmfchar*}(20,20)
\WthreeplainB
\fmfipair{w[]}
\fmfipair{wd[]}
\svertex{w4}{p4}
\svertex{w3}{p3}
\wigglywrap{w3}{v4}{v5}{w4}
\end{fmfchar*}}}
\;{\quad}&W_{B4}&=
\raisebox{\eqoff}{%
\fmfframe(3,1)(1,4){%
\begin{fmfchar*}(20,20)
\WthreeplainB
\fmfipair{w[]}
\fmfipair{wd[]}
\svertex{w6}{p6}
\svertex{w2}{p2}
\wigglywrap{w2}{v4}{v5}{w6}
\end{fmfchar*}}}
\\
W_{B5}&=
\raisebox{\eqoff}{%
\fmfframe(3,1)(1,4){%
\begin{fmfchar*}(20,20)
\WthreeplainB
\fmfipair{w[]}
\fmfipair{wd[]}
\svertex{w2}{p2}
\svertex{w5}{p5}
\wigglywrap{w2}{v4}{v5}{w5}
\end{fmfchar*}}}
\;{\quad}&W_{B6}&=
\raisebox{\eqoff}{%
\fmfframe(3,1)(1,4){%
\begin{fmfchar*}(20,20)
\WthreeplainB
\fmfipair{w[]}
\fmfipair{wd[]}
\svertex{w4}{p4}
\svertex{w2}{p2}
\wigglywrap{w2}{v4}{v5}{w4}
\end{fmfchar*}}}
\;{\quad}&W_{B7}&=
\raisebox{\eqoff}{%
\fmfframe(3,1)(1,4){%
\begin{fmfchar*}(20,20)
\WthreeplainB
\fmfipair{w[]}
\fmfipair{wd[]}
\svertex{w6}{p6}
\svertex{w1}{p1}
\wigglywrap{w1}{v4}{v5}{w6}
\end{fmfchar*}}}
\;{\quad}&W_{B8}&=
\raisebox{\eqoff}{%
\fmfframe(3,1)(1,4){%
\begin{fmfchar*}(20,20)
\WthreeplainB
\fmfipair{w[]}
\fmfipair{wd[]}
\svertex{w1}{p1}
\svertex{w5}{p5}
\wigglywrap{w1}{v4}{v5}{w5}
\end{fmfchar*}}}
\\
W_{B9}&=
\raisebox{\eqoff}{%
\fmfframe(3,1)(1,4){%
\begin{fmfchar*}(20,20)
\WthreeplainB
\fmfipair{w[]}
\fmfipair{wd[]}
\svertex{w1}{p1}
\svertex{w4}{p4}
\wigglywrap{w1}{v4}{v5}{w4}
\end{fmfchar*}}}
\end{aligned}
\end{equation*}
\caption{Wrapping diagrams with chiral structure $\dchi{2,1}$}
\label{diagrams12}
\end{figure}

\vspace{1cm}

\begin{table}[h]
\begin{center}
\fbox{
\small
\begin{tabular}{|m{35pt}m{35pt}|}
\hline
&\\[-2ex]
$W_{B1}\rightarrow$&$-W_{B2}$\\
$W_{B2}\rightarrow$&$-W_{B1}$\\
$W_{B3}\rightarrow$&finite\\
\hline
\end{tabular}
\begin{tabular}{|m{35pt}m{35pt}|}
\hline
&\\[-2ex]
$W_{B4}\rightarrow$&$-W_{B5}$\\
$W_{B5}\rightarrow$&$-W_{B4}$\\
$W_{B6}\rightarrow$&finite\\
\hline
\end{tabular}
\begin{tabular}{|m{35pt}m{35pt}|}
\hline
&\\[-2ex]
$W_{B7}\rightarrow$&$-W_{B8}$\\
$W_{B8}\rightarrow$&$-W_{B7}$\\
$W_{B9}\rightarrow$&finite\\
\hline
\end{tabular}}
\end{center}
\caption{Cancellations for diagrams with structure $\dchi{2,1}$}
\label{cancellations12}
\end{table}

\begin{figure}[h]
\unitlength=0.75mm
\settoheight{\eqoff}{$\times$}%
\setlength{\eqoff}{0.5\eqoff}%
\addtolength{\eqoff}{-12.5\unitlength}%
\settoheight{\eqofftwo}{$\times$}%
\setlength{\eqofftwo}{0.5\eqofftwo}%
\addtolength{\eqofftwo}{-7.5\unitlength}%
\begin{equation*}
\begin{aligned}
W_{C1\phantom{0}}&=
\raisebox{\eqoff}{%
\fmfframe(3,1)(1,4){%
\begin{fmfchar*}(20,20)
\WoneplainB
\fmfipair{w[]}
\fmfipair{wd[]}
\svertex{w5}{p5}
\svertex{w6}{p6}
\svertex{w3}{p3}
\vvertex{w7}{w6}{p3}
\fmfi{wiggly}{w6..w3}
\wigglywrap{w5}{v4}{v5}{w7}
\end{fmfchar*}}}
\;{\quad}&W_{C2\phantom{0}}&=
\raisebox{\eqoff}{%
\fmfframe(3,1)(1,4){%
\begin{fmfchar*}(20,20)
\WoneplainB
\fmfipair{w[]}
\fmfipair{wd[]}
\svertex{w5}{p5}
\svertex{w6}{p6}
\svertex{w3}{p3}
\vvertex{w7}{w6}{p3}
\fmfi{wiggly}{w6..w7}
\wigglywrap{w5}{v4}{v5}{w7}
\end{fmfchar*}}}
\;{\quad}&W_{C3\phantom{0}}&=
\raisebox{\eqoff}{%
\fmfframe(3,1)(1,4){%
\begin{fmfchar*}(20,20)
\WoneplainB
\fmfipair{w[]}
\fmfipair{wd[]}
\svertex{w4}{p4}
\svertex{w6}{p6}
\svertex{w3}{p3}
\vvertex{w7}{w6}{p3}
\fmfi{wiggly}{w6..w3}
\wigglywrap{w4}{v4}{v5}{w7}
\end{fmfchar*}}}
\;{\quad}&W_{C4\phantom{0}}&=
\raisebox{\eqoff}{%
\fmfframe(3,1)(1,4){%
\begin{fmfchar*}(20,20)
\WoneplainB
\fmfipair{w[]}
\fmfipair{wd[]}
\svertex{w4}{p4}
\svertex{w6}{p6}
\svertex{w3}{p3}
\vvertex{w7}{w6}{p3}
\fmfi{wiggly}{w6..w7}
\wigglywrap{w4}{v4}{v5}{w7}
\end{fmfchar*}}}
\\
W_{C5\phantom{0}}&=
\raisebox{\eqoff}{%
\fmfframe(3,1)(1,4){%
\begin{fmfchar*}(20,20)
\WoneplainB
\fmfipair{w[]}
\fmfipair{wd[]}
\svertex{w4}{p4}
\svertex{w6}{p6}
\svertex{w3}{p3}
\vvertex{w7}{w6}{p3}
\fmfi{wiggly}{w6..w7}
\wigglywrap{w4}{v4}{v5}{w3}
\end{fmfchar*}}}
\;{\quad}&W_{C6\phantom{0}}&=
\raisebox{\eqoff}{%
\fmfframe(3,1)(1,4){%
\begin{fmfchar*}(20,20)
\WoneplainB
\fmfipair{w[]}
\fmfipair{wd[]}
\svertex{w1}{p1}
\svertex{w6}{p6}
\svertex{w3}{p3}
\vvertex{w7}{w6}{p3}
\fmfi{wiggly}{w6..w3}
\wigglywrap{w1}{v4}{v5}{w7}
\end{fmfchar*}}}
\;{\quad}&W_{C7\phantom{0}}&=
\raisebox{\eqoff}{%
\fmfframe(3,1)(1,4){%
\begin{fmfchar*}(20,20)
\WoneplainB
\fmfipair{w[]}
\fmfipair{wd[]}
\svertex{w1}{p1}
\svertex{w6}{p6}
\svertex{w3}{p3}
\vvertex{w7}{w6}{p3}
\fmfi{wiggly}{w6..w7}
\wigglywrap{w1}{v4}{v5}{w7}
\end{fmfchar*}}}
\;{\quad}&W_{C8\phantom{0}}&=
\raisebox{\eqoff}{%
\fmfframe(3,1)(1,4){%
\begin{fmfchar*}(20,20)
\WoneplainB
\fmfipair{w[]}
\fmfipair{wd[]}
\svertex{w1}{p1}
\svertex{w6}{p6}
\svertex{w3}{p3}
\vvertex{w7}{w6}{p3}
\fmfi{wiggly}{w6..w7}
\wigglywrap{w1}{v4}{v5}{w3}
\end{fmfchar*}}}
\\
W_{C9\phantom{0}}&=
\raisebox{\eqoff}{%
\fmfframe(3,1)(1,4){%
\begin{fmfchar*}(20,20)
\WoneplainB
\fmfipair{w[]}
\fmfipair{wd[]}
\fmfipair{wu[]}
\svertex{w6}{p6}
\dvertex{wu4}{wd4}{p4}
\vvertex{w7}{w6}{p3}
\vvertex{w8}{wu4}{p3}
\fmfi{wiggly}{wu4..w8}
\wigglywrap{wd4}{v4}{v5}{w7}
\end{fmfchar*}}}
\;{\quad}&W_{C10}&=
\raisebox{\eqoff}{%
\fmfframe(3,1)(1,4){%
\begin{fmfchar*}(20,20)
\WoneplainB
\fmfipair{w[]}
\fmfipair{wd[]}
\fmfipair{wu[]}
\dvertex{wu4}{wd4}{p4}
\vvertex{w7}{wu4}{p3}
\fmfi{wiggly}{wu4..w7}
\wigglywrap{wd4}{v4}{v5}{w7}
\end{fmfchar*}}}
\;{\quad}&W_{C11}&=
\raisebox{\eqoff}{%
\fmfframe(3,1)(1,4){%
\begin{fmfchar*}(20,20)
\WoneplainB
\fmfipair{w[]}
\fmfipair{wd[]}
\fmfipair{wu[]}
\dvertex{wu4}{wd4}{p4}
\dvertex{wu3}{wd3}{p3}
\fmfi{wiggly}{wu4..wd3}
\wigglywrap{wd4}{v4}{v5}{wu3}
\end{fmfchar*}}}
\;{\quad}&W_{C12}&=
\raisebox{\eqoff}{%
\fmfframe(3,1)(1,4){%
\begin{fmfchar*}(20,20)
\WoneplainB
\fmfipair{w[]}
\fmfipair{wd[]}
\fmfipair{wu[]}
\svertex{w4}{p4}
\svertex{w6}{p6}
\vvertex{w3}{w4}{p3}
\vvertex{w7}{w6}{p3}
\fmfi{wiggly}{w4..w3}
\wigglywrap{w4}{v4}{v5}{w7}
\end{fmfchar*}}}
\\
W_{C13}&=
\raisebox{\eqoff}{%
\fmfframe(3,1)(1,4){%
\begin{fmfchar*}(20,20)
\WoneplainB
\fmfipair{w[]}
\fmfipair{wd[]}
\fmfipair{wu[]}
\svertex{w4}{p4}
\svertex{w3}{p3}
\fmfi{wiggly}{w4..w3}
\wigglywrap{w4}{v4}{v5}{w3}
\end{fmfchar*}}}
\;{\quad}&W_{C14}&=
\raisebox{\eqoff}{%
\fmfframe(3,1)(1,4){%
\begin{fmfchar*}(20,20)
\WoneplainB
\fmfipair{w[]}
\fmfipair{wd[]}
\fmfipair{wu[]}
\svertex{w4}{p4}
\svertex{w1}{p1}
\svertex{w6}{p6}
\vvertex{w7}{w4}{p3}
\vvertex{w8}{w6}{p3}
\fmfi{wiggly}{w4..w7}
\wigglywrap{w1}{v4}{v5}{w8}
\end{fmfchar*}}}
\;{\quad}&W_{C15}&=
\raisebox{\eqoff}{%
\fmfframe(3,1)(1,4){%
\begin{fmfchar*}(20,20)
\WoneplainB
\fmfipair{w[]}
\fmfipair{wd[]}
\fmfipair{wu[]}
\svertex{w4}{p4}
\svertex{w1}{p1}
\vvertex{w7}{w4}{p3}
\fmfi{wiggly}{w4..w7}
\wigglywrap{w1}{v4}{v5}{w7}
\end{fmfchar*}}}
\;{\quad}&W_{C16}&=
\raisebox{\eqoff}{%
\fmfframe(3,1)(1,4){%
\begin{fmfchar*}(20,20)
\WoneplainB
\fmfipair{w[]}
\fmfipair{wd[]}
\fmfipair{wu[]}
\svertex{w4}{p4}
\svertex{w1}{p1}
\svertex{w2}{p2}
\vvertex{w7}{w4}{p3}
\vvertex{w8}{w2}{p3}
\fmfi{wiggly}{w4..w7}
\wigglywrap{w1}{v4}{v5}{w8}
\end{fmfchar*}}}
\\
W_{C17}&=
\raisebox{\eqoff}{%
\fmfframe(3,1)(1,4){%
\begin{fmfchar*}(20,20)
\WoneplainB
\fmfipair{w[]}
\fmfipair{wd[]}
\fmfipair{wu[]}
\svertex{w1}{p1}
\svertex{w2}{p2}
\svertex{w4}{p4}
\vvertex{w7}{w2}{p3}
\vvertex{w8}{w4}{p3}
\fmfi{wiggly}{w2..w7}
\wigglywrap{w1}{v4}{v5}{w8}
\end{fmfchar*}}}
\;{\quad}&W_{C18}&=
\raisebox{\eqoff}{%
\fmfframe(3,1)(1,4){%
\begin{fmfchar*}(20,20)
\WoneplainB
\fmfipair{w[]}
\fmfipair{wd[]}
\fmfipair{wu[]}
\svertex{w1}{p1}
\svertex{w2}{p2}
\vvertex{w7}{w2}{p3}
\fmfi{wiggly}{w2..w7}
\wigglywrap{w1}{v4}{v5}{w7}
\end{fmfchar*}}}
\end{aligned}
\end{equation*}
\caption{Wrapping diagrams with chiral structure $\dchi{1}$}
\label{diagrams1}
\end{figure}


\begin{table}[h]
\begin{center}
\fbox{
\small
\begin{tabular}{|m{35pt}m{35pt}|}
\hline
&\\[-2ex]
$W_{C1\phantom{0}}\rightarrow$&$-W_{C5\phantom{0}}$\\
$W_{C2\phantom{0}}\rightarrow$&finite\\
$W_{C3\phantom{0}}\rightarrow$&$-W_{C11}$\\
$W_{C4\phantom{0}}\rightarrow$&$-W_{C10}$\\
$W_{C5\phantom{0}}\rightarrow$&$-W_{C1\phantom{0}}$\\
\hline
\end{tabular}
\begin{tabular}{|m{35pt}m{35pt}|}
\hline
&\\[-2ex]
$W_{C6\phantom{0}}\rightarrow$&finite\\
$W_{C7\phantom{0}}\rightarrow$&finite\\
$W_{C8\phantom{0}}\rightarrow$&finite\\
$W_{C9\phantom{0}}\rightarrow$&$-W_{C12}$\\
$W_{C10}\rightarrow$&$-W_{C4\phantom{0}}$\\
\hline
\end{tabular}
\begin{tabular}{|m{35pt}m{35pt}|}
\hline
&\\[-2ex]
$W_{C11}\rightarrow$&$-W_{C3\phantom{0}}$\\
$W_{C12}\rightarrow$&$-W_{C9\phantom{0}}$\\
$W_{C13}\rightarrow$&finite\\
$W_{C14}\rightarrow$&finite\\
$W_{C15}\rightarrow$&finite\\
\hline
\end{tabular}
\begin{tabular}{|m{35pt}m{35pt}|}
\hline
&\\[-2ex]
$W_{C16}\rightarrow$&finite\\
$W_{C17}\rightarrow$&finite\\
$W_{C18}\rightarrow$&finite\\
&\\
&\\
\hline
\end{tabular}}
\end{center}
\caption{Cancellations for diagrams with structure $\dchi{1}$}
\label{cancellations1}
\end{table}

\clearpage

\begin{table}[p]
\settoheight{\eqoff}{$\times$}%
\setlength{\eqoff}{0.5\eqoff}%
\addtolength{\eqoff}{-7.5\unitlength}
\begin{equation*}
\begin{aligned}
I_0=
\raisebox{\eqoff}{%
\begin{fmfchar*}(20,15)
\fmfleft{in}
\fmfright{out}
\fmf{plain}{in,v1}
\fmf{plain,left=0.25}{v1,v2}
\fmf{plain,left=0}{v2,v4}
\fmf{plain,tension=0.5,right=0.25}{v1,v0,v1}
\fmf{phantom,right=0.25}{v0,v3}
\fmf{plain}{v0,v2}
\fmf{plain}{v0,v4}
\fmffixed{(0.9w,0)}{v1,v3}
\fmffixed{(0.4w,0)}{v2,v4}
\fmfpoly{phantom}{v4,v2,v0}
\fmffreeze
\end{fmfchar*}}
&=\frac{1}{(4\pi)^6}\Big(
\frac{1}{6\varepsilon^3}-\frac{1}{2\varepsilon^2}
+\frac{2}{3\varepsilon}
\Big)
\\
I_1=
\raisebox{\eqoff}{%
\begin{fmfchar*}(20,15)
\fmfleft{in}
\fmfright{out}
\fmf{plain}{in,v1}
\fmf{plain,left=0.25}{v1,v2}
\fmf{plain,left=0}{v2,v4}
\fmf{plain,left=0.25}{v4,v3}
\fmf{plain,tension=0.5,right=0.25}{v1,v0,v1}
\fmf{plain,right=0.25}{v0,v3}
\fmf{plain}{v0,v2}
\fmf{plain}{v0,v4}
\fmf{plain}{v3,out}
\fmffixed{(0.9w,0)}{v1,v3}
\fmffixed{(0.4w,0)}{v2,v4}
\fmfpoly{phantom}{v4,v2,v0}
\fmffreeze
\end{fmfchar*}}
&=\frac{1}{(4\pi)^8}\Big(
-\frac{1}{24\varepsilon^4}+\frac{1}{4\varepsilon^3}
-\frac{19}{24\varepsilon^2}
+\frac{5}{4\varepsilon}
\Big)
\\
I_2=\raisebox{\eqoff}{%
\begin{fmfchar*}(20,15)
\fmfleft{in}
\fmfright{out}
\fmf{plain}{in,v1}
\fmf{plain,left=0.25}{v1,v2}
\fmf{plain,left=0.25}{v2,v3}
\fmf{plain,left=0.25}{v3,v4}
\fmf{plain,left=0.25}{v4,v1}
\fmf{plain,tension=0.5,right=0.5}{v2,v0,v2}
\fmf{phantom}{v0,v3}
\fmf{plain}{v1,v0}
\fmf{plain}{v0,v4}
\fmf{plain}{v3,out}
\fmffixed{(0.9w,0)}{v1,v3}
\fmffixed{(0,0.45w)}{v4,v2}
\fmffreeze
\end{fmfchar*}}
&=\frac{1}{(4\pi)^8}\Big(
-\frac{1}{24\varepsilon^4}+\frac{1}{4\varepsilon^3}
-\frac{19}{24\varepsilon^2}
+\frac{1}{\varepsilon}\Big(\frac{5}{4}-\zeta(3)\Big)
\Big)
\\
I_3=\raisebox{\eqoff}{%
\begin{fmfchar*}(20,15)
\fmfleft{in}
\fmfright{out}
\fmf{plain}{in,v1}
\fmf{plain,left=0.25}{v1,v2}
\fmf{plain,left=0.25}{v2,v3}
\fmf{plain,left=0.25}{v3,v4}
\fmf{plain,left=0.25}{v4,v1}
\fmf{plain,tension=0.5,right=0.25}{v1,v0,v1}
\fmf{phantom}{v0,v3}
\fmf{plain}{v2,v0}
\fmf{plain}{v0,v4}
\fmf{plain}{v3,out}
\fmffixed{(0.9w,0)}{v1,v3}
\fmffixed{(0,0.45w)}{v4,v2}
\fmffreeze
\end{fmfchar*}}
&=\frac{1}{(4\pi)^8}\Big(
-\frac{1}{12\varepsilon^4}+\frac{1}{3\varepsilon^3}
-\frac{5}{12\varepsilon^2}
-\frac{1}{\varepsilon}\Big(\frac{1}{2}-\zeta(3)\Big)\Big)
\\
I_4=\raisebox{\eqoff}{%
\begin{fmfchar*}(20,15)
\fmfleft{in}
\fmfright{out}
\fmf{plain}{in,v1}
\fmf{plain,left=0.25}{v1,v2}
\fmf{plain,left=0.25}{v2,v3}
\fmf{plain,left=0.25}{v3,v4}
\fmf{plain,left=0.25}{v4,v1}
\fmf{plain}{v2,v4}
\fmf{plain}{v3,v1}
\fmf{plain}{v3,out}
\fmffixed{(0.9w,0)}{v1,v3}
\fmffixed{(0,0.45w)}{v4,v2}
\fmffreeze
\end{fmfchar*}}
&=\frac{1}{(4\pi)^8}
\frac{1}{\varepsilon}5\zeta(5)
\\
I_5=\raisebox{\eqoff}{%
\begin{fmfchar*}(20,15)
\fmfleft{in}
\fmfright{out}
\fmf{plain}{in,v1}
\fmf{plain,tension=2,left=0.25}{v1,v2}
\fmf{plain,tension=2,left=0.25}{v2,v3}
\fmf{derplain,left=0.25}{v4,v1}
\fmf{plain,right=0.25}{v4,v0}
\fmf{plain,right=0}{v0,v1}
\fmf{plain,right=0.25}{v0,v5}
\fmf{plain,right=0.75}{v4,v5}
\fmf{phantom,right=0}{v3,v0}
\fmf{derplain,right=0.25}{v5,v3}
\fmf{plain}{v3,out}
\fmffixed{(0.9w,0)}{v1,v3}
\fmfpoly{phantom}{v2,v4,v5}
\fmffixed{(0.5w,0)}{v4,v5}
\fmf{plain,tension=0.5}{v2,v0}
\fmffreeze
\fmfshift{(0,0.15w)}{in,out,v1,v2,v3,v4,v5,v0}
\end{fmfchar*}}
&=
\frac{1}{(4\pi)^8}
\frac{1}{\varepsilon}(-\zeta(3))
\\
I_6=\raisebox{\eqoff}{%
\begin{fmfchar*}(20,15)
\fmfleft{in}
\fmfright{out}
\fmf{plain}{in,v1}
\fmf{derplain,tension=2,right=0.25}{v2,v1}
\fmf{derplain,tension=2,left=0.25}{v2,v3}
\fmf{plain,left=0.25}{v4,v1}
\fmf{plain,right=0.25}{v4,v0}
\fmf{plain,right=0}{v0,v1}
\fmf{plain,right=0.25}{v0,v5}
\fmf{plain,right=0.75}{v4,v5}
\fmf{phantom,right=0}{v3,v0}
\fmf{plain,right=0.25}{v5,v3}
\fmf{plain}{v3,out}
\fmffixed{(0.9w,0)}{v1,v3}
\fmfpoly{phantom}{v2,v4,v5}
\fmffixed{(0.5w,0)}{v4,v5}
\fmf{plain,tension=0.5}{v2,v0}
\fmffreeze
\fmfshift{(0,0.15w)}{in,out,v1,v2,v3,v4,v5,v0}
\end{fmfchar*}}
&=\frac{1}{(4\pi)^8}\frac{1}{\varepsilon}\Big(\frac{1}{2}\zeta(3)
-\frac{5}{2}\zeta(5)\Big)
\end{aligned}
\\
\end{equation*}
\caption{Momentum integrals}
\label{integrals}
\end{table}

\end{fmffile}

\clearpage

\footnotesize
\bibliographystyle{JHEP}
\bibliography{references-def}

\end{document}